\documentclass[floatfix,superscriptaddress,a4paper,showpacs,showkeys,nofootinbib,notitlepage,12pt]{revtex4-1}
\usepackage[colorlinks=true,linktocpage=true,linkcolor=blue,citecolor=blue,allcolors=blue]{hyperref}
\usepackage{epsfig}
\usepackage{latexsym}
\usepackage[utf8]{inputenc}
\usepackage{xspace}
\usepackage{indentfirst}
\usepackage{enumitem}
\usepackage{color}
\usepackage{placeins}

\usepackage{setspace}
\usepackage{lipsum}

\usepackage{hyperref}

\usepackage{todonotes}

\usepackage{amsmath}
\usepackage{amssymb}
\usepackage[english]{babel}
\usepackage{url}
\graphicspath{{figs/}}
\newcommand{\hspm}{\hspace*{0.5pt}}

\newcommand{\eq}[1]{\begin{align} #1 \end{align}}

\newcommand{\be}{\begin{equation}}
\newcommand{\ee}{\end{equation}}

\begin{document}

\title{
Pion stars  embedded in neutrino clouds}

\author{O. S. Stashko}
\affiliation{\mbox{Princeton Center for Theoretical Science, Princeton University, Princeton, NJ 08544}}
\affiliation{\mbox{Goethe Universität, Max-von-Laue Str. 1, Frankfurt am Main, 60438, Germany}}

\author{O. V. Savchuk}
\affiliation{Facility for Rare Isotope Beams, Michigan State University, East Lansing, MI 48824 USA}
\affiliation{Bogolyubov Institute for Theoretical Physics, 03680 Kyiv, Ukraine}

\author{L.~M. Satarov}
\affiliation{\mbox{Frankfurt Institute for Advanced Studies, D-60438 Frankfurt am Main, Germany}}

\author{I.~N.~Mishustin}
\affiliation{\mbox{Frankfurt Institute for Advanced Studies, D-60438 Frankfurt am Main, Germany}}

\author{M.~I. Gorenstein}
  \affiliation{Bogolyubov Institute for Theoretical Physics, 03680 Kyiv, Ukraine}
  \affiliation{\mbox{Frankfurt Institute for Advanced Studies, D-60438 Frankfurt am Main, Germany}}
  
\author{V.~I. Zhdanov}
\affiliation{Taras Shevchenko National University of Kyiv, 03022 Kyiv, Ukraine}

\date{\today}

\begin{abstract}
We study  self-gravitating multi-pion systems (pion stars) in a state of the Bose condensate.
To ensure stability of such stars, it is assumed that they are immersed in the lepton  background.   Two different  phenomenological equations of state (EoS) for the pion matter are used, some of them having the first order phase transition.

The model  parameters are chosen to 
reproduce the recent lattice QCD data at zero temperature and large isospin chemical potential.
It is shown that the mass-radius diagrams of pion stars obtained with phenomenological EoS are close to ones calculated in the ideal gas model. We analyze properties of neutrino clouds which are necessary for stabilizing the pion stars.
\end{abstract}
\maketitle

\section{Introduction}

The cores of astrophysical objects can have sufficiently high  densities, at which the nuclear EoS or even the hadron-quark  phase transition \cite{PhysRevLett.105.161102,PhysRevC.85.032801, PhysRevC.89.025806} may have observable signatures. Multimessenger astronomy provides important constrains on the properties of strongly interacting matter.  In particular, recent observations of gravitational waves from neutron star mergers are used for constraining theoretical models for  EoS of~stellar matter\cite{Dietrich_2020,PhysRevC.100.055801,
Hanauske2020a,Hanauske2020b,Schaffner2018,
PhysRevD.104.063003, Drischler_2021,Huth_2022,https://doi.org/10.48550/arxiv.2208.00994}.  New data from  LIGO-VIRGO-KAGRA detectors are expected to observe new neutron-star mergers \cite{Baibhav_2019,Colombo_2022,Patricelli_2022}.  New capabilities are associated with the future launch of the LISA mission \cite{LISAconsort} to provide additional  constraints on the nuclear EoS from data of neutron star masses in binary systems. The Dune   and Hyper-Kamiokande  neutrino observatories under construction   will be able to provide data on the physics of supernova explosions and the physics of neutrinos \cite{DUNE, Kamiokande,  Mezzetto}. The prospects of observational technology stimulate interest in exotic astrophysical configurations that can be considered as possible alternatives to the black holes.

Considerable attention is paid to boson star models~\cite{Schunck_2003,2017Liebling,Braaten_2016}, where  the  Bose-Einstein condensation (BEC) in astrophysical objects is discussed. Systems with BEC were considered in    \cite{Su_rez_2013,Bernal_2017,Visinelli_2016,HajiSadeghi_2019}   as candidates for the dark matter.
Unlike these articles,  there were attempts to consider BEC within the Standard Model and study models of  astrophysical objects made of pions \cite{Brandt:2018bwq,Mannarelli_2019,andersen2018boseeinstein}.   Following \cite{Brandt:2018bwq}, these objects will be called below as pion stars (PS).   
At small temperatures 
these stars contain the  Bose condensate of charged pions.
The pion  condensates have been widely discussed for decades in astrophysics in connection with neutron stars (see, e.g., \cite{Migdal, Umeda1994} and references therein), they could also be formed in the early Universe 
\cite{Vovchenko} and can appear in heavy ion collisions~\cite{Begun:2006gj,Begun:2008hq}.

The EoS at low temperature and nonzero isospin chemical potential  
was recently studied by lattice Quantum Chromodynamics (lQCD) simulations~\cite{Brandt:2017oyy,Brandt:2018bwq,Vovchenko:2020crk,Brandt_2022a}.  Pions are expected to be the dominant degrees of freedom at such conditions. In particular, the Bose-Einstein condensation   at isospin chemical potentials close to the pion mass has been observed. The results of these first-principle simulations were used to estimate properties of~PS  in~Ref.~\cite{Brandt:2018bwq}. 
However, characteristics of the  external 
neutrino cloud, necessary for the PS stability, were not considered by these 
authors.

Different models were used to calculate the phase diagram of the pion matter, see, e.g., Refs.~\cite{Adhikari:2019zaj,Adhikari:2020kdn,He:2005nk,Adhikari:2018cea,Folkestad:2018psc}.
Recently, the pion matter EoS was
considered within the effective mass model 
\cite{Mishustin:2019otg,Anchishkin_2019,Savchuk:2020yxc,Mishustin:2019otg}
and in the mean-field model 
\cite{stashko2020thermodynamic,PhysRevC.104.055202,PhysRevC.106.034319}.
These two phenomenological models are used in the present paper.
Their parameters are chosen to reproduce the lQCD data from Ref.~\cite{Brandt_2022a}. Both versions, with and without a first order phase transition 
(FOPT), are considered. One of our goals is to check  whether a presence of the FOPT  will change the PS properties, in particular, its mass-radius diagram. 
Also, we pay special attention to stability of PS with respect to weak decays of pions taking into account that this  can be achieved  if the star is embedded in the neutrino cloud of galactic 
size~(cf.~\cite{Brandt:2018bwq}).

The paper is organized as follows. In Sec.~\ref{EoS} the phenomenological models for the pion matter EoS and stability conditions in PS are considered. In Sec.~\ref{comp} the contributions of the pion and lepton components of the PS are discussed. Section \ref{sec-TOV} presents calculations of mass-radius diagram for the PS. In Sec.~\ref{sec-ic} we consider properties of the neutrino cloud surrounding the inner core of PS,
and a short summary in Sec.~\ref{sum} closes the paper.

\section{EoS of isospin-asymmetric pion matter}
\label{EoS}

\subsection{Ideal gas model}
Below we consider isospin-asymmetric pion systems at zero temperature and nonzero isospin chemical potential $\mu$. If interactions are neglected, all pions are at rest and form   
the pion  Bose condensate where 
$\mu=m_\pi$  (here  $m_\pi\simeq 140~\textrm{MeV}$ is the pion mass). 

Within the ideal gas model the pion pressure vanishes, and the energy density \mbox{$\varepsilon=m_{\pi}\, |n|$}, where $n=n_{\pi^+}-n_{\pi^-}$ is the pion isospin density. 
In a stable macroscopic~PS,
the Coulomb interactions
and weak pion decays should be suppressed. This can be achieved by including  
charged leptons $e$ and $\mu$ as well as neutrinos $\nu_e$ and~$\nu_\mu$~\cite{Brandt:2018bwq}. The number densities of charged leptons $n_l$, pressure~$p_l$, and energy densities~$\varepsilon_l$ are the functions of the corresponding chemical potentials $\mu_l$. They are  determined by well-known formulae of the ideal relativistic Fermi gas
($\hbar=c=1$):
\eq{ & n^{\rm id}_{l}(\mu_l)= \frac{g_l}{6\pi^2} \, (\mu_{l}^{2}- m_{l}^2)^{3/2}\theta (\mu_{l}-m_{l}),
\label{Fermi-dens} \\
& p^{\rm id}_{l}=\int_{0}^{\mu_l}n^{\rm id}_{l}(\mu)\hspace*{1pt}d\mu,\\ & \varepsilon^{\rm id}_{l}=\mu_l n^{\rm id}_{l}-p^{\rm id}_{l},
\label{Fermi-dens1}
}
where $l=(e,\mu)$, $g_l=2$, and $\theta(x)$ is a theta function.
We take the mass values: 
\mbox{$m_\mu=105.6~\rm{MeV}$} and \mbox{$m_e=0.511~\rm{MeV}$}.
The same 
expressions (\ref{Fermi-dens})--(\ref{Fermi-dens1}) are valid for massless left-handed neutrinos 
after replacing
$l\to\nu_l$,~$g_l\to 1$, and $m_l\to 0$. 

\subsection{Effective mass model}
Now we introduce the interaction effects, regarding pions as the only interacting component of the PS. These effects are modelled with two phenomenological EoS.
First, we consider the effective mass (EM) model. It was formulated in Ref.~\cite{Mishustin:2019otg} for the pion system at zero chemical potential
and later applied for interacting alpha particles  in Ref.~\cite{Satarov:2020loq}. 
Within the EM model pions are represented by 
a~triplet of the interacting scalar fields
$\phi=(\phi_1,\phi_2,\phi_3)$ with the effective Lagrangian density
\eq{
{\mathcal{L}}=\frac{1}{2} \left(\partial_\mu\phi\,
\partial^{\,\mu}\phi
- m_\pi^2\hspm\phi^2\right)+\mathcal{L}_{\rm int}\left(\phi^2\right)\,,
\label{lagr1}
}
where $\mathcal{L}_{\rm int}$ is the interaction part of the Lagrangian.
In the mean-field approximation one can represent $\mathcal{L}_{\rm int}$
as a series over the powers of $\delta\sigma=\phi^2-\sigma$, where
\mbox{$\sigma=<\phi^2>$} is the average scalar density of pions in the grand canonical ensemble.
Taking into account only the lowest-order terms, one arrives at the mean-field
Lagrangian~\cite{Mishustin:2019otg}
\eq{
 {\mathcal{L}}\approx
\frac{1}{2}\left[\partial_\mu\phi\,\partial^{\,\mu}\phi
- M^2(\sigma)\hspace*{1pt}\phi^2\right]+p_{\rm ex}(\sigma)\,,
\label{mfa}
}
where $M(\sigma)$ is the effective pion mass and $p_{\rm ex}(\sigma)$ is the so-called excess pressure,
\eq{
M^2(\sigma)=m_\pi^2-2\hspace*{1pt}\frac{d\mathcal{L}_{\rm int}}{d\sigma}\,,~~p_{\rm ex}(\sigma)=\mathcal{L}_{\rm int}\hspace*{.5pt} (\sigma)-
\sigma\hspace*{1pt}\frac{d\mathcal{L}_{\rm int}}{d\sigma}\,.
\label{mfa1}
}
Following Ref.~\cite{Mishustin:2019otg}, we use a Skyrme-like parametrization of $\mathcal{L}_{\rm int}$:
\eq{
\label{lis}
\mathcal{L}_{\rm int}(\sigma)=\dfrac{a}{4}\hspm\sigma^2-\frac{b}{6}\hspm\sigma^3\,,
}
where $a$ and $b$ are the model parameters which describe, respectively,
attractive (at $a>0$) and repulsive ($b>0$)  interactions between (quasi)particles. 
At~$a=0$ and $b=0$ one gets a~limiting case of the pion ideal gas model.
Substituting (\ref{lis}) into (\ref{mfa1}) one obtains the following expressions for $M$ and  $p_{\rm ex}$:
\eq{
M(\sigma)=\sqrt{m_{\pi}^2-a \sigma +b \sigma^2},\,~
p_{\,\rm ex}(\sigma)=
-\frac{a}{4} \sigma^2 +\frac{b}{3} \sigma^3\,.\label{epre2}
}

Within the considered model,  FOPT may occur in the pionic matter in the case of non-zero positive $a$. At $T=0$ this transition takes place between the vacuum~(g) and the condensed (liquid) phase (l). These two phases corresponds to zeros of $p_{\rm ex}$, namely, to scalar densities $\sigma=\sigma_g = 0$ and $\sigma=\sigma_l =3a/4b$.
The binding energy per pion in the condensed phase is nonzero and equals $W=M_l-m_\pi$.

\begin{figure}[htb!]
    \centering
    \includegraphics[width=0.49\textwidth]{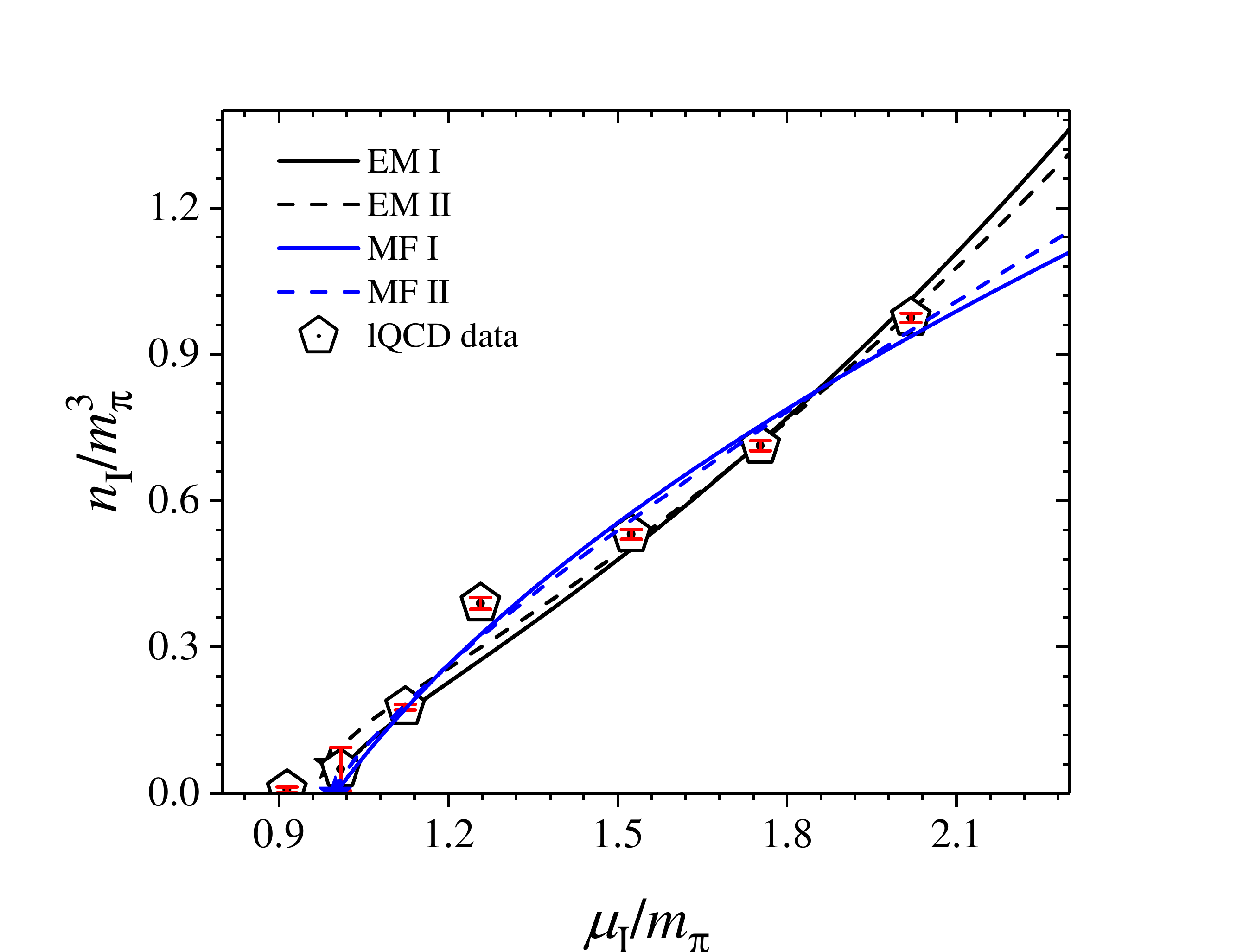}
    \includegraphics[width=0.49\textwidth]{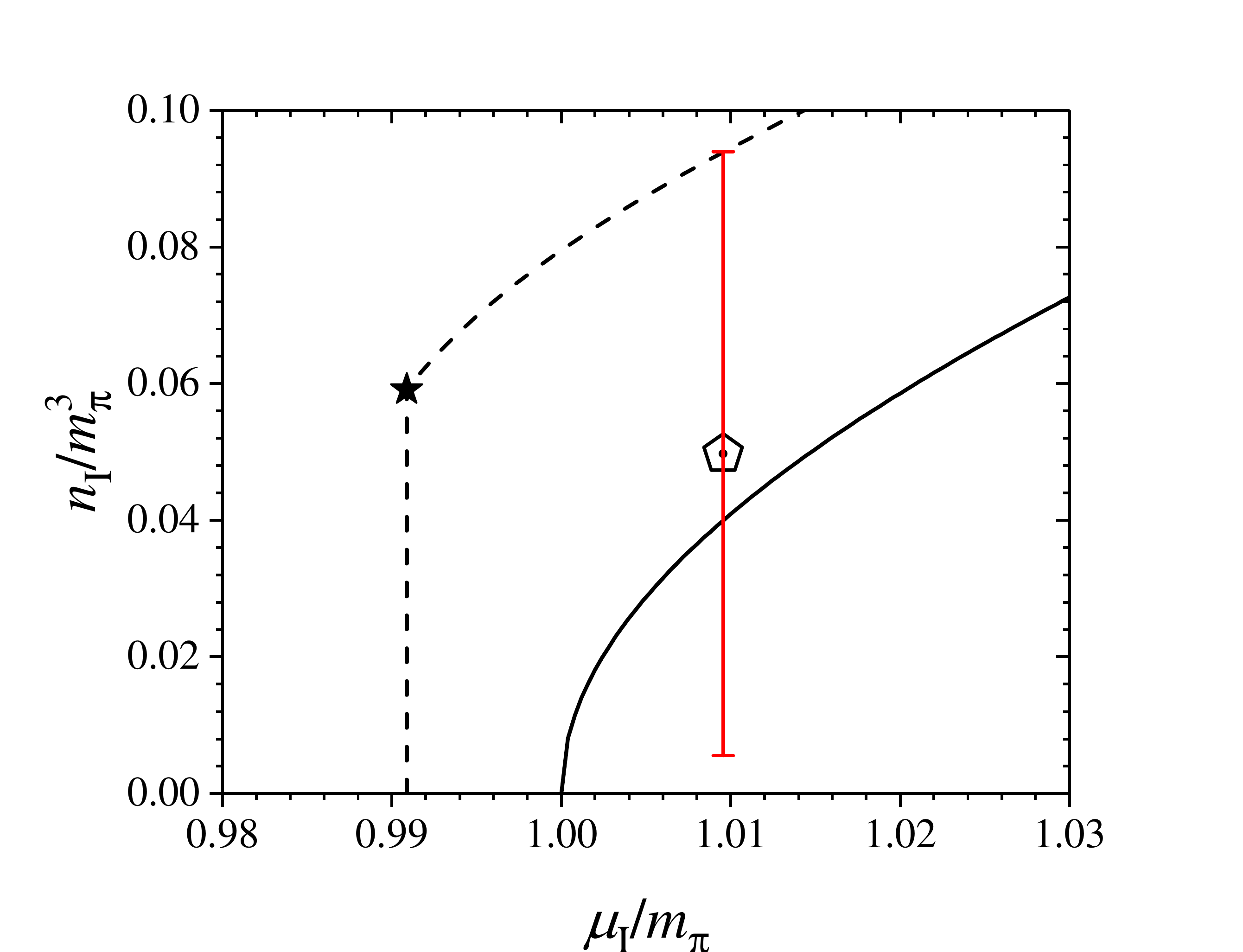}
    \caption{A comparison of the model results for the pion isospin density $n_I$ as a function of the isospin chemical potential $\mu_I$ at 
    $T=0$ with the lattice data of Ref.~\cite{Brandt_2022a}.
    Right panel shows enlarged part of
    the left lower corner.
    }
    \label{fig:fit}
\end{figure}

 Using Eq.~(\ref{epre2}), one gets the following values of the effective mass in the vacuum and in the condensed phases:
\eq{\label{sM}
M_g=m_{\pi}\,,~~~~M_l=\sqrt{m_{\pi}^2-\frac{3a^2}{16b}}\hspm .
}
The BEC of cold equilibrium pionic matter occurs at the isospin 
pion chemical potential 
$\mu=M_l$. 
To calculate the pion  density $n$ as the function of $\mu$ one should solve the system of equations $n=\mu\hspm\sigma,~ \mu=M(\sigma)$. 
The parameters $a$ and $b$ are fitted to the lQCD data~\cite{Brandt_2022a}.
The best fit is denoted by EM\,II. The corresponding parameters are shown in Table~I. The quality of the fit is demonstrated in Fig.~\ref{fig:fit}. To investigate the sensitivity to the FOPT, we present also the results for the purely repulsive pion interaction with $a=0$ (set EM\,I). As expected, both FOPT and the bound state of pion matter do not appear in this case (see last two columns of Table~I).
\begin{table}[h!]
\begin{tabular}{|c|c|c|c|c|}
\hline
    & $a$ & $b~[{\rm MeV^{-2}}]$&~FOPT~&$W~{\rm [MeV]}$\\ \hline

EM I    & $0$  &  $6.2\cdot10^{-4}$   & \textrm{absent} & $0$\\ \hline
EM II  & $1.22$  &  $7.8\cdot10^{-4}$  & \textrm{exists} &$-1.28$ \\ \hline
\end{tabular}
\caption{The values of interaction parameters in the EM model. }
\end{table}

\FloatBarrier

\subsection{Mean field model}
The second phenomenological model used in this paper is the mean field~(MF) model. This model introduces a density-dependent mean-field  potential $U(n)$ which shifts the pion chemical potential $\mu$ with respect to its  ideal gas value. 
The pion mass is fixed to its vacuum value. Assuming the Skyrme-like parametrization of $U(n)$, 
\eq{
U(n)~=-~A\, n~+~B\, n^{\gamma+1}~,\label{Un}
}
one can write the equation for the pion number density $n$ as a function of the chemical potential $\mu$. 
At zero temperature, when all pions are in the Bose condensed state, it  has the following form
~\cite{PhysRevC.104.055202,PhysRevC.106.034319}
\eq{\mu-U(n)=m_\pi~. }
The pressure $p\hspace*{1pt}(\mu)$  is found after integrating 
the pion density over $\mu$.

The parameters of the mean-field potential $A,B$, and~$\gamma$ are again found from the best fit of the lQCD data~\cite{Brandt_2022a}. We consider the cases of a soft ($\gamma=1/3$) and hard (\mbox{$\gamma=1$}) repulsion. In the first case the lattice data are better reproduced with positive $A$ (attraction). However, for $\gamma=1$ a purely repulsive potential
is preferable (see Table~II).
FOPT exists  for \mbox{$A>0$}. The parameters of this transition 
are found by finding nontrivial solutions of the equa\-tion $p\hspace*{1pt}(\mu)=0$.
The coefficients of the Skyrme interaction are listed in Table~II. 
Comparison of the EM~I, EM~II, MF~I, and MF~II models with the lattice data is presented in  Fig.~\ref{fig:fit}~(a). As will be seen later, the most important region of the pion EoS for the PS structure is  $\mu\approx m_\pi$. This region is shown separately in Fig.~\ref{fig:fit} (b) for the EM~I and EM~II models. 
\begin{table}[h!]
\begin{tabular}{|c|c|c|c|c|}
\hline
      & {\footnotesize $A~[{\rm MeV\cdot fm^3}]$} & {\footnotesize $B~[{\rm MeV\cdot fm^{3(\gamma+1)}}]$} & {\footnotesize FOPT} & {\footnotesize $W~{\rm [keV]}$} \\ \hline
MF I  &$-246.81$   & $536.4$ &~$\textrm{absent}$~& 0 \\ \hline
MF II & $224.03$   & $772.36$&~$\textrm{exists}$~&$-6.1$\\ \hline
\end{tabular}
\caption{The values of interaction parameters in the MF model.}
\end{table}

\section{Full EoS for the pion star matter}
\subsection*{A. Equilibrium conditions}
Stable PS
can not consist of pions only. Indeed, the electric charge of (positively) charged pions  must be compensated by (negatively) charged leptons ($\mu$ and/or $e$).   
The charged pions in the vacuum 
undergo the weak 
decays. In particular, 
$\pi^+\to\mu^+ + \nu_\mu$ 
proceeds with the
lifetime of about $2.6 \times 10^{-8}\hspm\textrm{s}$.
Two other decay modes are $\pi^+\to e^++\nu_e$ and
$\pi^+\to\pi^0+e^++\nu_e$.
Muons also decay in the vacuum, e.g., via $\mu^-\rightarrow e^-+\overline{\nu}_e+\nu_\mu$. 
In~stable PS 
the above decays should be suppressed. This suppression can be provided by the Pauli blocking of neutrinos.
Thus, in addition to $\mu^-$ and $e^-$ which neutralize electric charge, one needs also
both $\nu_\mu$ and $\nu_e$. Therefore, a minimal set of particles in the~PS is $(\pi^+,\mu^-,e^-,\nu_\mu,\nu_e)$ or, equivalently, $(\pi^-,\mu^+,e^+,\overline \nu_\mu,\overline \nu_e)$. We denote this set as $\pi l \nu$.

 In our calculations we impose the following constraints: the local charge neutrality, 
 \eq{Q=n_{I} +n_{\mu}+n_{e}~=0~,\label{el}
 }
and the chemical equilibrium between all constituents: 
\eq{
\mu_I & =\mu_{\mu^+}+\mu_{\nu_\mu}=
\mu_{e^+}+\mu_{\nu_e}~\nonumber \\
& = - \mu_{\mu^-}+\mu_{\nu_\mu}=
-\mu_{e^-}+\mu_{\nu_e} ~.\label{chem}
}

At given $n_{I}$ and $\mu_I$, the lepton chemical potentials
and corresponding number densities are found from the conditions (\ref{el}) and (\ref{chem}),
and the ideal gas equation (\ref{Fermi-dens}).
Similar to
Ref.~\cite{Brandt:2018bwq}, we  assume the equality $\mu_{\nu_e}=\mu_{\nu_\mu}$, which is motivated by neutrino oscillations.
Note that the truncated sets of particle species, like $\pi$, $\pi l$, or $\pi \nu$,  do not satisfy the stability conditions in the PS.

In the case of the FOPT, the transition between the vacuum and the pion Bose condensate takes place at some value of the chemical potential $\mu= \mu_{\rm BC}$ in the interval of pion densities 
$n<n_{BC}$. To implement  FOPT into calculations we apply the mixed phase construction similar to the Maxwell construction for the liquid-gas phase transition. At \mbox{$T=0$} both liquid and gas phases of the phase transition have vanishing pressure and this makes the pionic EoS with the FOPT softer.

\subsection*{B. Components of the pion star}\label{comp}
\begin{figure*}[bht!]
\includegraphics[width=\textwidth]{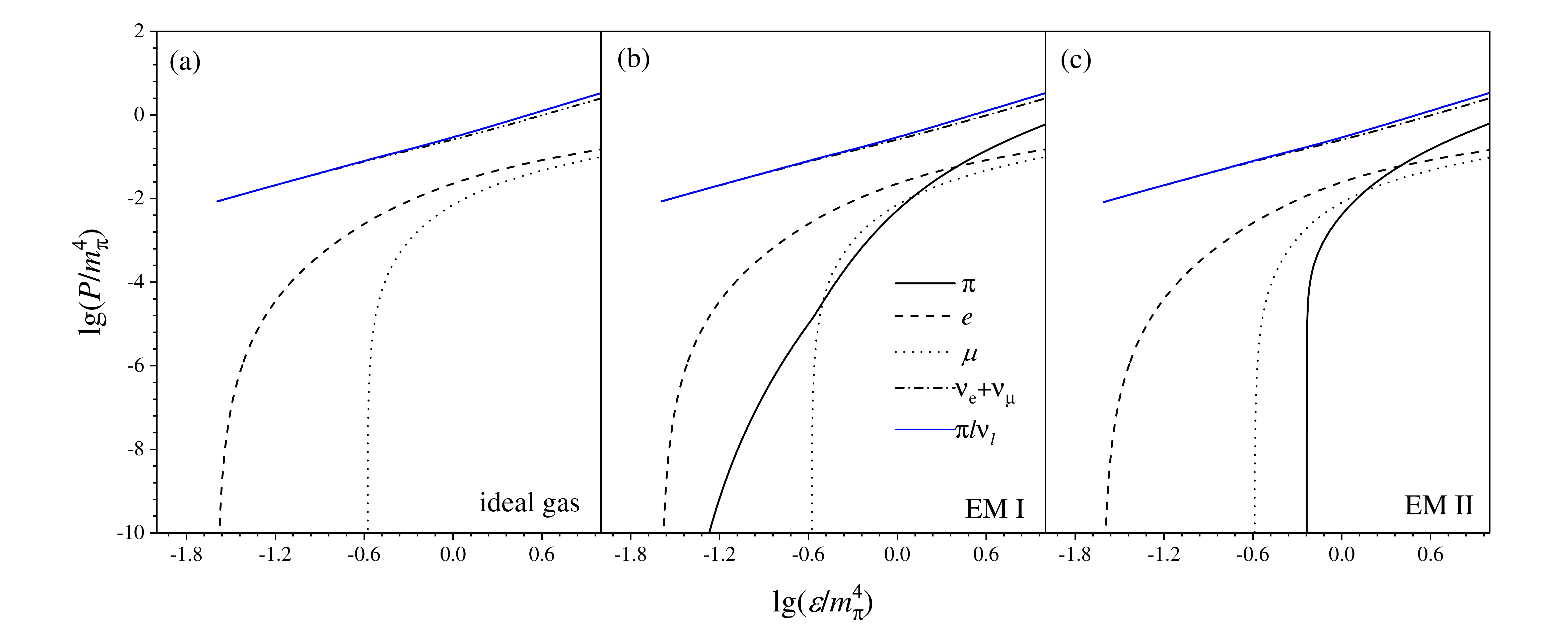}
    \caption{\label{p-e} Partial pressures $p_i$ of different particles species and the total pressure $P$ as functions of the total energy density $\varepsilon$ for the ideal gas (a), EM~I (b)  and EM~II (c) models of the pion EoS. }
    \label{fig:pe}
\end{figure*}
As mentioned above, we treat leptonic components of the PS as mixture of ideal Fermi gases. Their thermodynamic
 functions are given by Eqs.~(\ref{Fermi-dens})--(\ref{Fermi-dens1}) 
 with the conditions of electro-neutrality (\ref{el}) and chemical equilibrium (\ref{chem}).
 
The complete EoS for the PS matter is then defined 
by the following set of equations: 
\eq{
& n_I= n_\mu+n_e~,~~~~P=p_{\pi} +p_\mu +p_e+p_{\nu_\mu}+p_{\nu_e}~, \label{pe-1}\\
& \varepsilon=\varepsilon_{\pi}+\varepsilon_\mu+
\varepsilon_e + \varepsilon_{\nu_\mu}+\varepsilon_{\nu_e} ~.\label{pe-2}
}
 Figures \ref{p-e} show the partial contributions $p_i$ 
 of different system components to the total pressure $P$  as functions of the total energy density $\varepsilon$
 for the ideal gas (a), EM~I (b), and EM~II~(c) models of the pion EoS.   
 For all three EoS one has the relations \mbox{$p_\mu<p_e<p_\nu\approx P$}. Note that within the ideal gas model the pion  pressure vanishes. 
 However, the pion pressure $p$ becomes non-zero in the system of interacting pions, and it changes with $\varepsilon$ differently in the EM~I and EM~II models. 
 Nevertheless, the relation $p\ll p_\nu \approx P$ holds in these models. 
 As a~consequence, the sensitivity of $P=P(\varepsilon)$  to the pion matter EoS  
 is very weak. Therefore, it is no wonder that EM and MF  models are only slightly deviate  from the ideal gas of pions where $p=0$. 
In all considered cases the neutrino pressure provides the main contribution  as compared to other constituents. 
By this reason, 
 the results for the PS radial profiles are rather  robust (see the next section).
On the other hand, our calculations show that the size of the pion core for the $\pi l \nu$ star exhibits some sensitivity to the presence of the FOPT.

\section{TOV equations and mass-radius diagrams}\label{sec-TOV}

The structure of a static spherical star composed of an ideal isotropic fluid  is described by the well-known Tolman-Oppenheimer-Volkoff (TOV) equations which can be written in the form \cite{Tolman_1939,Oppen_1939}
\eq{\frac{dM}{dr}&=4\pi r^2 \varepsilon(r)~,
\nonumber \\ \frac{dP}{dr}&=-G~\frac{[\varepsilon(r)+P(r)][M(r)+4\pi r^3 P(r)]}{r[r-2G M(r)]}~,
\label{TOV}}
where $G$ is the gravitation constant, and $M(r),~\varepsilon(r)$, and $P(r)$  are, respectively, the integrated mass, energy density and pressure of the fluid
at the distance $r$ from the star center. 

We solve these equations numerically at given EoS 
$P=P (\varepsilon)$ 
defined by Eqs.~(\ref{pe-1}) and (\ref{pe-2}).
The boundary condition assumes $M(0)=0$ and  a certain value of the central
pressure~$P(0)$.
By integrating the Eq.~(\ref{TOV}) with  
the conditions at $r\to 0$:
\eq{
M = \frac{4}{3}\pi r^3\varepsilon(0)+
O(r^4),~P= P(\varepsilon(0))+O(r),~
} 
one obtains the radial profiles $\varepsilon (r)$. The integration is done outward, up to the  
surface \mbox{$r=R_*$}\hspm, where the densities of pions and charged leptons vanish and  neutrinos remain the only matter components. The external neutrino cloud may extend to infinity (see the next section). This is a general feature for the EoS of massless particles.

\begin{figure*}[hbt!]
    \centering
    \includegraphics[width=0.49\textwidth]{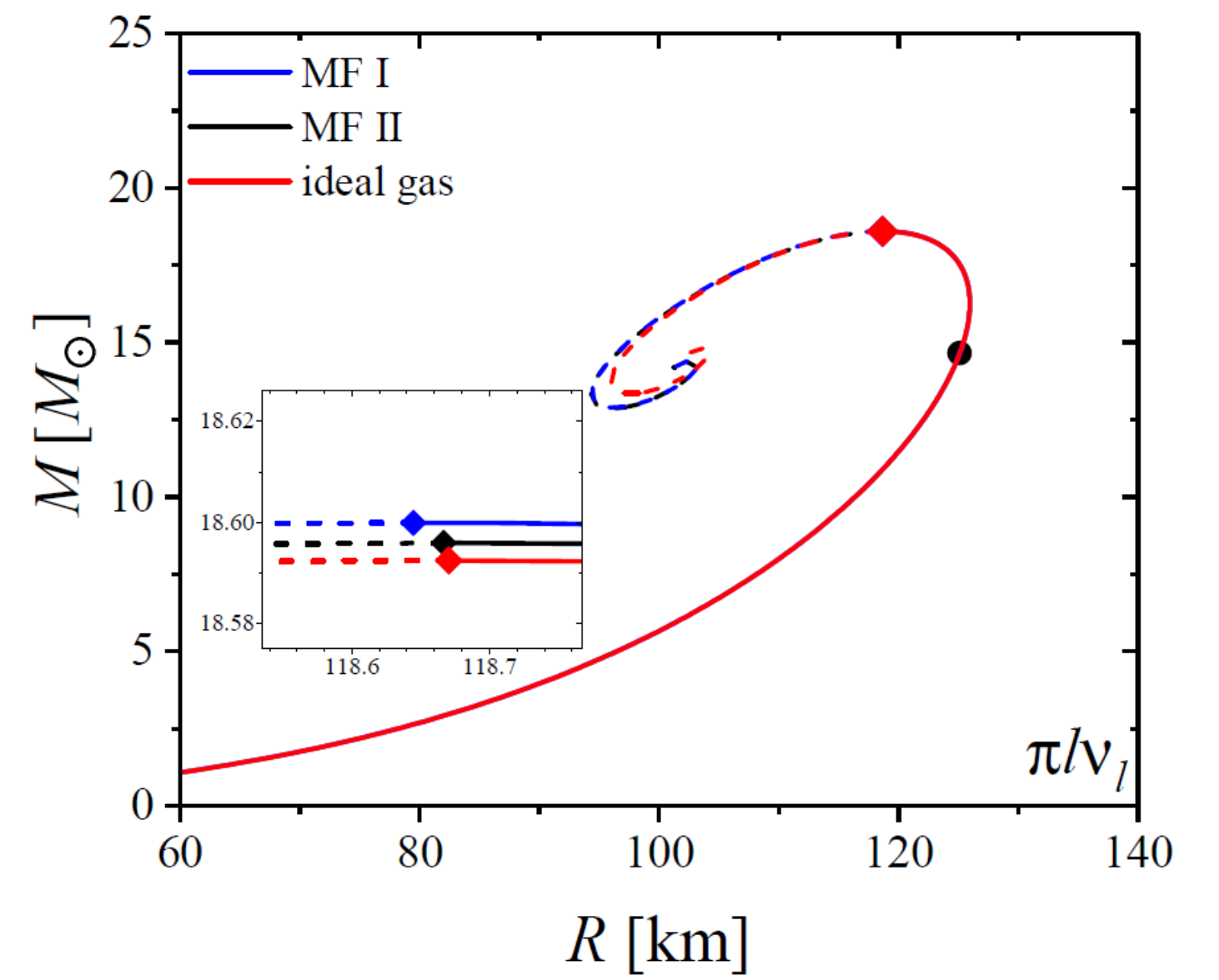}
    \includegraphics[width=0.49\textwidth]{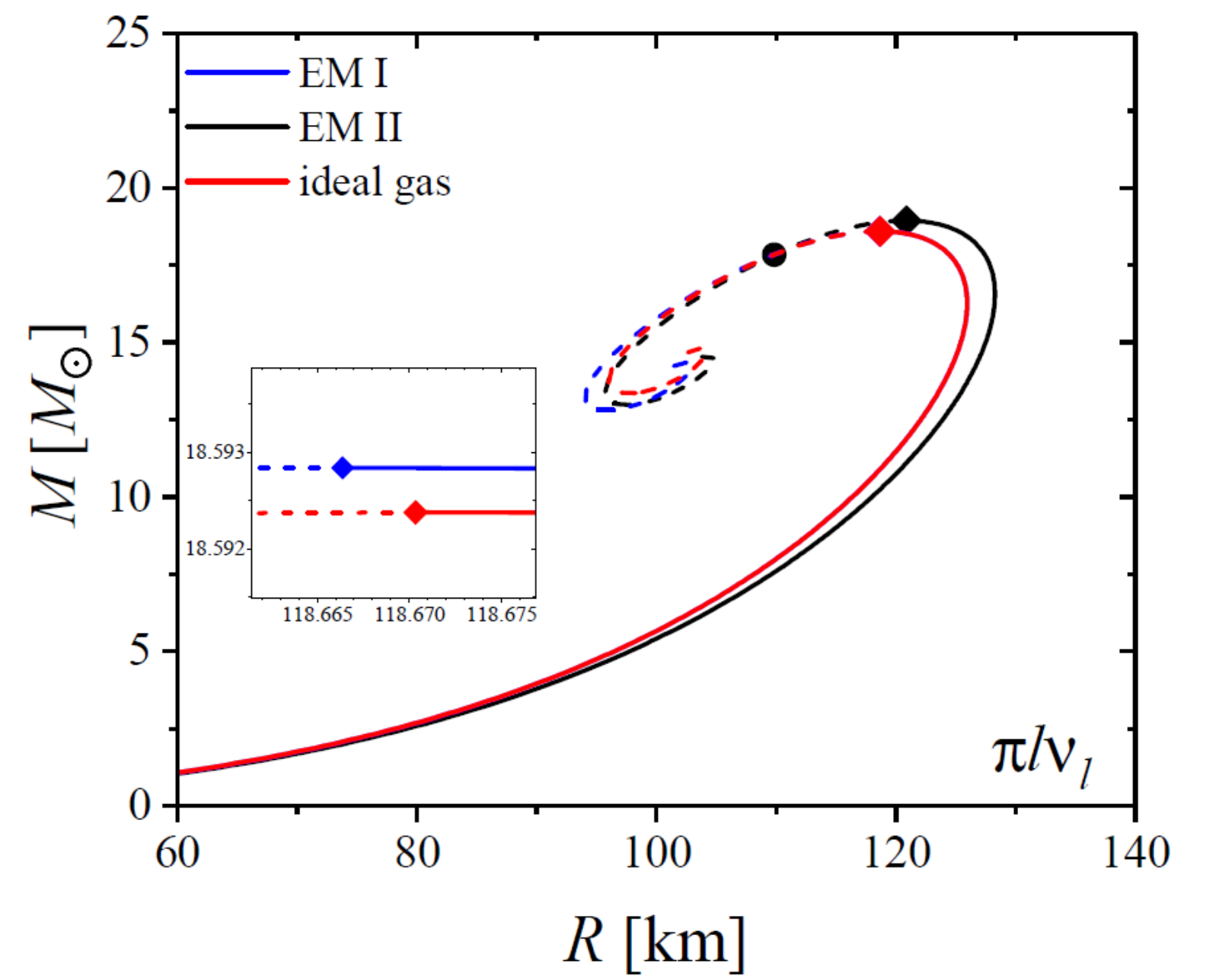}
    \caption{Mass-radius diagrams of the PS within the MF (left) and EM (right) models. Dashed parts of diagrams correspond to unstable configurations. Inserts represent the results near the states of the maximal mass (shown by diamonds). The black circles represent the points of the phase transition.}
    \label{fig:MR}
\end{figure*}
We define the mass of the PS as $M(R_{*})=M_{*}$.   Physical properties of the star, e.g., the radial profile of pressure, can be found through their dependence on $\varepsilon(r)$.
By considering an~ensemble of initial $\varepsilon_c=\varepsilon(0)$ we obtain the line on the $M_{*}-R_{*}$ plane. This can be done for different EoS 
of the pion matter. The mass-radius (MR) diagrams for the PS are presented in Fig.~\ref{fig:MR}.

As seen from Fig.~\ref{fig:MR} 
the ideal pion gas and interacting pion models lead to very similar MR diagrams with maximal masses \mbox{$M_*\approx 19~M_{\odot}$} and corresponding  radii \mbox{$R_*=(120\pm 1)~\textrm{km}$}\hspm. The EM~II  model exhibits some sensitivity to the phase transition. As compared to EM~I, in this case one gets {smaller pion and neutrino} pressure, and obtains  larger radius ($R_*\approx 121\,\textrm{km}$).
Thus, there is some sensitivity 
of the MR
diagram to the FOPT in the pion matter.
The neutrino cloud exists  beyond the pionic core of the PS 
at $r>R_*$ (see the next section).
We checked that implementing non-zero masses of neutrinos leads to finite sizes of neutrino cloud. 
However, properties of the PS at distances  $r<R_*$ are not much influenced by this modification at small enough neutrino masses.

\section{Properties of stellar neutrinosphere }\label{sec-ic}

In this section we discuss properties of the neutrino cloud
surrounding the PS inner core with radius $R_*$\hspm\footnote
{
It is obvious that the 'naked' PS 
(without an external neutrino cloud) is 
unstable due to nonzero neutrino flux through the surface $r=R_*$. This in turn leads to the decays of pions at $r<R_*$\hspm. In such a process the PS can partially or totally evaporate.
} . 
We do not discuss dynamical processes leading to formation of
PS and assume that the radial structure of 
the final static star (the core + cloud) is given by the solution of the TOV 
equations~(\ref{TOV}). 

As mentioned above,
we define the boundary of the inner core as the radius where
densities of pions and charged leptons vanish. 
Similar to \cite{Brandt:2018bwq}, our calculations show that muons do not appear [\hspm $\mu\hspm (r)<m_\pi+m_\mu$]\hspm\footnote
{
In this section we omit index $\nu$\hspm. 
} in the PS even for states with maximal core mass $M_*$\hspm.
In accordance with the chemical equilibrium
condition, the boundary chemical potential of neutrino equals $\mu_*=m_\pi+m_e\simeq 140.5~\textrm{MeV}$.
Corresponding number- and energy densities of neutrino (both flavours)
at zero temperature can be calculated {as}
\eq{\label{bvd}
n_*=\dfrac{\mu_*^3}{3\pi^2}\simeq 0.0122~\textrm{fm}^{-3},~~
\varepsilon_*=\dfrac{\mu_*^4}{4\pi^2}\simeq 1.29~\textrm{MeV}\cdot \textrm{fm}^{-3}.
}
Note that these densities can be regarded as the threshold values, above which the pion condensate can be formed inside the cold neutrino matter.

Below we consider the
state with maximal mass of inner core, $M_*\equiv M(R_*)$ and disregard pion interactions.
In this case, the numerical solution of the TOV equations gives \mbox{$R_*\simeq 120~\textrm{km}$}, \mbox{$M_*\simeq 19\, M_{\odot}$}\,. The calculation shows that the values $\mu_*$ and $n_*$ are, respectively, about 64\% and 26\%
of the corresponding central values.

Using the TOV equations, we have checked that at $r>R_*$ the energy density decreases (approximately) inversely proportional to $r^2$:
\eq{\label{edr}
\varepsilon=\frac{\mu^4}{4\pi^2}\simeq\varepsilon_*\left(\dfrac{R_*}{r}\right)^2.
}
Formally, this corresponds to the linear increase of the neutrino cloud mass with values
\mbox{$M-M_*\propto (r-R_*)$}.
However, these relations should be modified at large $r$ due to nonzero neutrino rest mass~$m$. Deviations
 from ultrarelativistic approximation $\mu\gg m$ occur above some maximal radius, $r\gtrsim r_{\rm max}$\hspm . The latter can be estimated by substituting $\mu=m$ into Eq.~(\ref{edr}). Then one obtains
\eq{\label{ema}
r_{\rm max}\sim R_*\left(\dfrac{\mu_*}{m}\right)^2.
}

One can consider this radius as a size of the neutrino cloud. 
Choosing $m=1~\textrm{eV}$ (this does not contradict current observations) 
one obtains the estimate \mbox{$r_{\rm max}\sim 2.4\cdot 10^{18}$}~$\textrm{km}\simeq 2.5\cdot 10^5~\textrm{ly}$  and mass $M \sim 10^{17}M_{\odot}$.  
The size $r_{\rm max}$ is comparable with the size of the dark matter halo of our Galaxy~\cite{Dea20}, but the mass (cf. also \cite{Narain}) is of the order of that for the largest known matter concentrations in the Universe.

One should have in mind, that these estimates do not take into account that the TOV equations~(\ref{TOV})
are not justified at the dilute periphery of the neutrino cloud. Indeed, the local thermodynamic equilibrium, used in derivation of these equations, breaks down when the local
Knudsen number, $\textrm{Kn}(r)=\lambda\hspm (r)/r$ becomes larger than 
unity\hspm\footnote
{
In particular, at ${\rm Kn}\gtrsim 1$, the pressure tensor of the neutrino cloud becomes anisotropic with
different transverse and radial components. Calculating the energy density profiles at such distances requires a~dedicated kinetic approach. Presumably, nonequilibrium (dissipation) effects will modify the asymptotic behavior of $\varepsilon\hspm (r)$ as compared to (\ref{edr}). At larger radii the neutrino cloud can be described
only within a kinetic approach. 
Nevertheless, we expect  that the TOV equations can still be used 
for rough  order-of-magnitude estimates. 
}.
Here $\lambda\hspm (r)$ is the mean-free path of neutrino  at the distance $r$ from the PS center:
\eq{\label{mfp}
\lambda\hspm (r)=\dfrac{1.6}{n<\hspm\sigma\hspm>},
} 
where $n=n_{\nu_e}+n_{\nu_\mu}$ is the total neutrino density, and $<\sigma>$ is the cross section
of the $\nu_e\nu_e$ scatterings averaged over the momentum distributions of neutrinos. 
Coefficient~$1.6$ takes into account that the $\nu_e\nu_\mu$ cross section is by a factor of 4 
smaller~\cite{Flo76} than that for~$\nu_e\nu_e$ (at~the same center of mass  energy squared $s$ of the neutrino pair). 

Using the explicit expression for $\sigma (s)$, given in Ref.~\cite{Flo76}, after averaging over the local momentum distribution of neutrinos, one gets the relations
\begin{eqnarray}
&&<\sigma>=\dfrac{1}{\pi}G_F^2<s>=\dfrac{9}{8\pi}\left(G_F\mu\right)^2=
\sigma_*\left(\dfrac{n}{n_*}\right)^{2/3},\label{sav1}\\
&&\sigma_*=\dfrac{9}{8\pi}\left(G_F\mu_*\right)^2\simeq 3.71\cdot 10^{-40}~\textrm{cm}^2,\label{sav2}
\end{eqnarray}
where $G_F\simeq 2.3\cdot 10^{-22}~\textrm{cm/MeV}$ is the Fermi coupling constant. 
Substituting (\ref{sav1}) into (\ref{mfp}) gives 
\eq{\label{mfp1}
\lambda=\lambda_*\left(\dfrac{n_*}{n}\right)^{5/3}\sim\lambda_*\left(\dfrac{r}{R_*}\right)^{5/2},
}
where $\lambda_*=1.6/(n_*\sigma_*)\simeq 3.53~\text{m}$. Note that a much smaller neutrino mean free path, about $37~\textrm{cm}$, is obtained at the PS center. The resulting Knudsen number
\mbox{$\textrm{Kn}\sim\textrm{Kn}_*(r/R_*)^{3/2}$}, where $\textrm{Kn}_*\simeq\lambda_*/R_*\sim 3.0\cdot 10^{-5}$. This shows that profiles of energy density, predicted by TOV equations,
are reliable only up to the radii \mbox{$r\lesssim R_*/\textrm{Kn}_*^{2/3}\sim$} ~$10^3\hspm R_*\simeq 1.25\cdot 10^5~\textrm{km}$.  

\section{Summary}\label{sum}
In the present paper we have considered pion stars defined as the self-gravitating configurations with the pion  Bose condensate.
 The local electric neutrality requires additional constituents to be present along the charged pion condensate. 
 Therefore, electrons and muons  are added. However, this is not sufficient to have chemical equilibrium,
 and both muon and electron neutrinos have to be added as well. The mass-radius diagrams of the PS are calculated by solving the TOV equations for different phenomenological pion EoS.
 Our calculations show that charged leptons and neutrinos contribute significantly to the pressure and 
energy density.
This makes the results for the interacting pions  to be almost identical to those of the ideal pion gas. 
 This finding provides a robust prediction of the PS mass-radius diagram. 
Some sensitivity to  FOPT still 
remains in size of the PS inner core.

 Whether PS can be considered as realistic astrophysical objects is still an open question. 
 In fact, if we want to limit ourselves to a model of a stationary configuration consisting of pions, leptons and massive neutrinos, we can make ends meet. However, unrealistically large mass within the galactic dimensions makes the existence of such object doubtful. To formulate an astrophysically relevant model, one must either include additional components of nuclear matter (as inside neutron stars), or consider a highly non-stationary configuration, or work outside the Standard Model.

\section*{Acknowledgements}
The authors are thankful to Scott Pratt and Horst Stoecker for fruitful comments and discussions. O.S.S. is thankful to Horst Stoecker for the warm hospitality at Frankfurt Institute for Advanced Studies. O.V.S. acknowledges support by the Department of Energy, Office of Science through grant no. DE-FG02-03ER41259. This work is supported by the National Academy of Sciences of Ukraine, Grant No.  0122U200259. L.M.S. and I.N.M.
thank for the support from the Frankfurt Institute for Advanced
Studies.
 M.I.G. acknowledges the support from the Alexander von Humboldt Foundation. V.I.Z. acknowledges partial support from the National Research Foundation of Ukraine (Project No. 2020.02/0073). This work was supported by a grant from the Simons Foundation (Grant Number 1039151).

\bibliography{main.bib}
\end{document}